\documentclass[doublecol]{epl2} 

\title{Toward a microscopic description of flow near the jamming \\threshold}

\author{E. Lerner\inst{1} \and G. D\"uring\inst{1} \and M. Wyart\inst{1}}

\institute{                    
  \inst{1} New York University, Center of Soft Matter Research - 4 Washington Place, New York, NY 10003, USA
}

\abstract{We study the relationship between microscopic structure and viscosity in 
non-Brownian suspensions.
We argue that the  formation and opening of contacts between particles in flow effectively leads to a negative
selection of  the contacts carrying weak forces.
We show that an analytically tractable model capturing this negative selection correctly reproduces scaling properties 
of flows near the jamming transition.
In particular, we predict that $(i)$ the viscosity $\eta$ diverges with the coordination number
$z$ as $\eta\sim (z_c-z)^{-(3+\theta)/(1+\theta)}$, $(ii)$ the operator which governs flow
displays  a low-frequency mode that controls the divergence of viscosity, at a frequency
$\omega_{\rm min}\sim (z_c-z)^{(3+\theta)/(2+2\theta)}$, and $(iii)$ the distribution of forces 
displays a  scale $f^*$ that vanishes near jamming as
$f^*/\langle f\rangle \sim(z_c-z)^{1/(1+\theta)}$ where $\theta$ characterizes 
the distribution of contact forces $P(f)\sim f^\theta$ at jamming, and 
where $z_c$ is the Maxwell threshold for rigidity.}

\begin{document}

\maketitle

Suspensions are heterogeneous fluids containing solid particles, whose viscosity
was computed early on by Einstein \cite{Einstein2} and Batchelor \cite{Batchelor} in the dilute regime.
As the packing fraction $\phi$ is increased however, steric hindrance becomes dominant and particles
move under stress in an increasingly coordinated way \cite{pouliquen2004,olsson,durian,heussinger2010a}. For non-Brownian particles, the viscosity $\eta$ eventually diverges as the suspension
jams into an amorphous solid at some packing fraction $\phi_c$. Recently, progress has been made in characterizing the rheological properties in this limit.
It has been shown that non-Brownian  suspensions \cite{boyer, lespiat, lemaitreroux}, as well as 
aerial granular flows \cite{dacruz,jop}, are characterized by two  constitutive relations  that relate the packing fraction $\phi$ and the macroscopic friction $\mu$ to the ratio of the local shear rate to the pressure. Both constitutive relations display singularities as jamming is approached,  in particular it is observed that $\eta\sim(\phi_c-\phi)^{-a}$ where $a\approx 2$ \cite{boyer}.
This phenomenological description is  valid beyond a length scale that grows near jamming, and below which non-local effects play a role \cite{pouliquen2004,lespiat,staron}. There is currently no accepted microscopic description for this growing length scale, nor for the observed constitutive relations. 

Olsson and Teitel \cite{olsson,teitel2011} and others \cite{heussinger2009,hatano08a} have popularized a simplified 
model of non-Brownian suspensions where hydrodynamical interactions are neglected. 
We refer to the hard-particle limit of this model as the Affine Solvent Model (ASM). The ASM constitutive relations 
are very similar to those found in real suspensions \cite{edan}, supporting that it captures the 
essential physics near jamming.  In this model it was observed that \cite{edan}: $(i)$~there exists a scaling 
relation between the viscosity $\eta$ and the coordination $z$: $\eta \sim (z_c-z)^{-2.85}$,  where $z_c=2D$ 
\cite{footnote} and $D$ is the spatial dimension. $(ii)$~The dynamics is governed by one operator only.
The material can thus be characterized  by the spectrum of this operator, which contains more information than the constitutive relations.  In flow, the spectrum displays bi-scaling near jamming, with a single mode being responsible for 
the fast divergence of the viscosity. In this Letter we explain these observations, make a new 
scaling prediction on the distribution of contact forces that we confirm empirically, and define and
measure two new exponents characterizing contact forces both in flow and at jamming.

ASM  is fully defined by the following three assumptions:  
$(i)$~hydrodynamic  interactions are neglected: the viscous drag on a particle 
is proportional to the difference ${\vec V_{\rm n.a.}}$
between the particle velocity and  the imposed velocity of the underlying 
fluid: ${\vec F}=-\xi_0 {\vec V_{\rm n.a.}}$ where $\xi_0$ is a drag coefficient. The flow 
of the fluid phase is undisturbed by the particles and is chosen to be an affine simple shear of strain
rate $\dot\gamma$. $(ii)$~The dynamics is over-damped. 
$(iii)$~Particles are hard, i.e. cannot overlap, and frictionless. 

As is also the case for non-Brownian suspensions of hard particles \cite{boyer,lemaitreroux}, within ASM 
 rheological properties depend on only
one dimensionless parameter, the normalized pressure 
$ p\equiv  p_p d^{D-2}/\dot\gamma\xi_0$ \cite{edan}. Here $ p_p$ is the particle
pressure and $d$ is the particle size.
Near jamming the particle shear stress $\sigma_p$ and pressure $p_p$ are proportional
$\mu\equiv\sigma_p/ p_p\!\rightarrow\!\mu^*\!>\!0$ \cite{peyneau,xu2006,teitel2011},
implying that the viscosity $\eta$ and the renormalized pressure $p$ are 
proportional too:  $\eta\equiv  \sigma_p/\dot\gamma\rightarrow(\xi_0\mu^*/d^{D-2})\times p$.
These quantities depend only on the geometry of the network formed by particles 
in contact and can be expressed in a compact form \cite{edan}: 
\begin{eqnarray}
\label{1}
\eta &=&\!\! -\frac{\langle \gamma|f\rangle}{\dot{\gamma}\Omega D} =  \frac{\xi_0}{\Omega} \langle \gamma|{\cal N}^{-1} |\gamma\rangle, \\ 
p_p &=& \frac{\langle r|f\rangle}{\Omega D} =  -\frac{\dot{\gamma}\xi_0}{\Omega D}\langle r | {\cal N}^{-1}|\gamma \rangle,
\end{eqnarray}
where $\Omega$ is the volume of the system. $|r\rangle$ is the vector of dimension $N_c$ -- the total number of contacts made between the $N$ particles -- of the distances $r_\alpha$ between particles in contact. 
$|\gamma\rangle$ is a vector, also of dimension $N_c$,  whose
components are the variations of the contact lengths under an affine simple shear in the $(x,y)$ plane:
$\gamma_\alpha\equiv \partial r_\alpha/\partial\gamma\equiv ({\vec r}_\alpha\!\cdot{\vec e}_x)( {\vec r}_\alpha\!\cdot{\vec e}_y)/r_\alpha$,
where ${\vec r}_\alpha$ is the vector between the centers of the particles forming the contact~$\alpha$. 
${\cal N}$ is a symmetric operator of dimension $N_c\times N_c$, which can be
written as ${\cal N}\equiv {\cal T}^t {\cal T}$.  ${\cal T}$ is the operator of dimension
$ND\times N_c$ that assigns to any set of $N_c$ contact forces $|f\rangle$ the associated net unbalanced forces $|F\rangle$ appearing 
on the particles \cite{calladine}:
\begin{equation}
\label{2}
|F\rangle ={\cal T}|f\rangle \ \ \Leftrightarrow \ \  \forall i, \ \ {\vec F}_{i}=\sum_{\alpha_i} {\vec n}_{\alpha_i} f_{\alpha_i},
\end{equation}
where $\alpha_i$ labels the contacts made by particle $i$ and where ${\vec n}_{\alpha}={\vec r}_\alpha/r_\alpha$. The non-zero elements of
${\cal T}$ thus correspond to the unit vectors ${\vec n}_{\alpha}$. ${\cal T}^t$ is its transpose.

Previously, we have numerically performed \cite{edan} a spectral analysis of 
${\cal N}\equiv\sum_\omega \omega^2 |f_\omega\rangle\langle f_\omega|$ in flow and found 
that: $(i)$~the spectrum of ${\cal N}$ displays bi-scaling and consists of two structures: one isolated 
mode of frequency $\omega_{\rm min}$, and a plateau of modes appearing above a 
frequency $\omega^*\sim p^{-\delta}$ with 
$\delta \approx 0.35$, as shown in Fig.2{\bf c},{\bf d}. $(ii)$~The divergence of the 
viscosity is governed by the lowest frequency modes $|f_{\rm min}\rangle$, which has a finite 
projection on the shear direction
\begin{equation}
\lim_{N\rightarrow\infty}\langle f_{\rm min}
|\gamma\rangle/(||f_{\rm min}||\times ||\gamma||)=O(1)>0 \ .
\end{equation}
According to Eq.(\ref{1}) this observation implies that $\eta\sim p\sim1/\omega_{\rm min}^2$, as shown 
in Fig.3{\bf c}.

\begin{figure}
\begin{center}
\onefigure[width=0.38\textwidth]{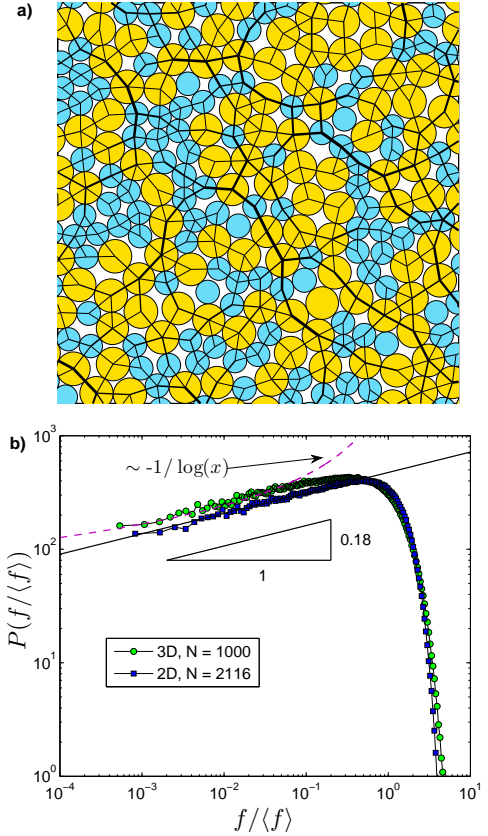}
\caption{{\bf a)} Snapshot of a shear-jammed configuration in $2D$. The thickness of the lines are proportional to the contact forces, the components
of $|\tilde f_0\rangle$, see text. {\bf b)} Distributions of the contact forces in shear-jammed configurations; the dashed curve is a fit of the $3D$ distribution to the functional form $P(x)\sim -1/\log(x)$,
and the black line is a power-law with exponent $\theta=0.18$. }
\label{f1}
\end{center}
\end{figure}

In this Letter we show that 
these  observations, together with the dependence of viscosity on coordination, can be explained  by one assumption only, namely 
that the configurations with 
coordination $z$ visited by the dynamics are similar to shear-jammed configurations (of friction $\mu^*$ and of coordination $z_c$)
where contacts carrying the smallest forces are removed until the coordination is $z$.  
The term ``similarity" is used here to indicate that the rheological properties of the two models, i.e. the real 
configurations found in flow and the constructed ones, fall in the same universality class. The 
rational for this similarity is that during flow, a negative selection of the weak contacts occur: indeed 
only contacts with a small force are fragile, i.e.~tend to open and disappear as flow progresses. On the 
other hand, new contacts are formed by collision and can immediately carry a significant force. Our 
rule to generate configurations is the simplest one that captures such a negative selection and can 
be analyzed analytically.  

\begin{figure*}
\begin{center}
\onefigure[width=0.8\textwidth]{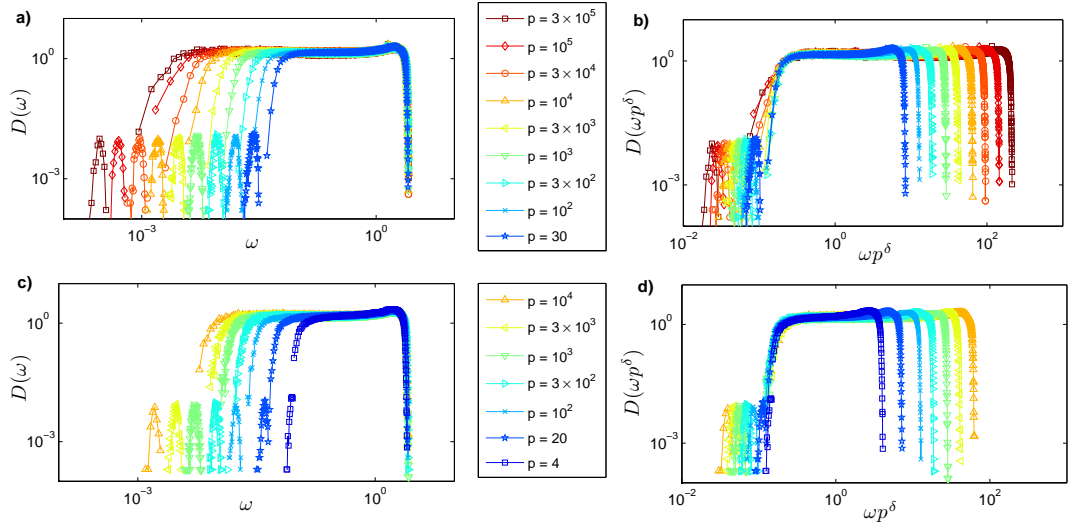}
\caption{
{\bf a)}~Average spectra $D(\omega)$ of the operator ${\cal N}$  {\it v.s.} frequency $\omega$ measured for configurations constructed from shear-jammed 
configurations by gradually removing contacts carrying the weakest forces, see text for details. The spectra are measured by binning the constructed configurations
according to their various pressures $p$ as indicated in the legend. 
The peak at low frequency  corresponds to one mode, and has been amplified by some factor for visibility.
{\bf b)}~$D(\omega)$ {\it v.s.} rescaled frequency $\omega p^\delta$ with $\delta = 0.35$. Note the collapse of the onset of the plateau of modes. 
{\bf c)}~ $D(\omega)$ {\it v.s.} frequency $\omega$ of the operator ${\cal N}$, measured for the configurations generated in simulations of flow \cite{edan}. 
For each pressure, the spectra are averaged over at least 300 independent flow configurations. 
{\bf d)}~$D(\omega)$ {\it v.s.} rescaled frequency $\omega p^\delta$ with $\delta = 0.35$ \cite{edan}.
}
\label{f2}
\end{center}
\end{figure*}

To check our assumption, we use an event-driven code  \cite{simulation_paper} to simulate flow. 
We simulate systems under simple shear flow with $N=1000$ particles in three dimensions, using Lees-Edwards periodic
boudary conditions \cite{ATbook}, at the volume fraction $\phi = 0.644 < \phi_c$. Half of the particles are small and half are large; we set the diameter ratio
of small and large particles to be 1.4. At large densities jamming can occur spontaneously 
when the coordination fluctuates up to $z=z_c$ (see \cite{simulation_paper} for details),
generating anisotropic configurations with an effective friction $\mu=\mu^*$.
We consider shear-jammed configurations that jam after
a shear strain of at least 200\% is imposed on an isotropic system. 
After averaging over 300 shear-jammed states
we obtain an average friction coefficient $\langle\mu^*\rangle = 0.125$, 
with a standard deviation $\sqrt{ \langle (\mu^* - \langle\mu^*\rangle)^2\rangle}  = 0.017$,
indicating that finite size fluctuations are small.  This result is consistent with previous observations \cite{selfAveragingMu} showing that $\mu^*$ is a self-averaging quantity,
whose standard deviation decreases as $1/\sqrt{N}$.


At jamming $\eta=\infty$ and  according to Eq.(\ref{1}) there must be one normalized 
mode $|\tilde f_0\rangle$ such that $\tilde{\cal N}|\tilde f_0\rangle=0$. 
Henceforth we use the tilde notation to refer to quantities characterizing shear-jammed configurations.
Thus
\begin{equation}
\label{3}
 0=\langle\tilde f_0|\tilde{\cal N}|\tilde f_0\rangle= \langle\tilde f_0|\tilde{\cal T}^t\tilde{\cal T}|\tilde f_0\rangle=||\tilde{\cal T} |\tilde f_0\rangle||^2,
\end{equation}
implying, together with the definition of $\tilde{\cal T}$, that $|\tilde f_0\rangle$ is the vector of 
contact forces that maintain force-balance. An example of $|\tilde f_0\rangle$ is shown in Fig.(\ref{f1}), together with the 
distribution of the contact forces computed over 300 shear-jammed configurations. 
The distribution of low forces is of particular importance,
and we find that $P(f/\langle f\rangle) \sim (f/\langle f\rangle)^\theta$ with  $\theta \approx 0.18$.
This scaling relation holds well for two decades for $D=2$. Such small exponents 
are hard to distinguish from the case where $\theta=0$ with logarithmic correction,
although the power law fits data better, especially for $D=2$, see Fig.(\ref{f1}).

We next apply the following procedure to each three-dimensional shear-jammed state: contact forces are sorted,
and the  $m\equiv \tilde N_c(z_c-z)/z_c$  weakest contacts are removed from the contact network, starting from $m=1$. Physically, removing a contact corresponds 
to eroding the particles at the  contact point, so that a finite gap appears. Mathematically, one simply does not include these contacts when 
computing the operator ${\cal T}$ defined in Eq.(\ref{2}).
For each $m$ we recompute the associated matrix ${\cal N}\equiv {\cal T}^t{\cal T}$ of dimension $N_c\times N_c$ with $N_c= \tilde N_c-m$.

\begin{figure}
\begin{center}
\onefigure[width=0.49\textwidth]{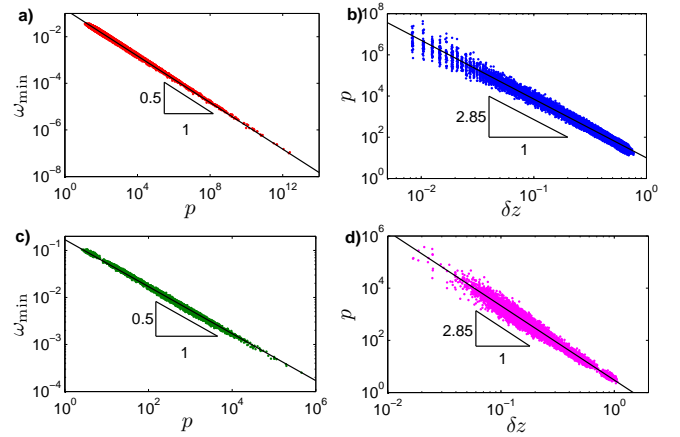}
\caption{Minimal frequency $\omega_{\rm min}$ of the operator ${\cal N}$ for our constructed configurations {\bf a)} and for configurations 
from simulations of flow {\bf c)}.
Pressure $p$ vs.~distance to threshold coordination $\delta z = z_c - z$ for constructed configurations
{\bf b)} and for flow {\bf d)}. These data are well-captured by the relations $\omega_{\rm min} \sim 1/\sqrt{p}$ and $p \sim \delta z^{-1/\delta}$ with $\delta \approx 0.35$
for both ensembles of configurations.}
\label{f3}
\end{center}
\end{figure}

We now argue that these constructed configurations are accurate models of configurations actually visited in flow. 
We first compare the spectra of ${\cal N}$ in the two cases and show that they scale similarly with pressure.   
Numerical diagonalization of ${\cal N}$ readily gives the density of states $D(\omega)$, displayed  in Fig.\ref{f2}{\bf a} for our constructed configurations
and in Fig.\ref{f2}{\bf c} for configurations from flow simulations. These quantities are indeed nearly identical, since both ($i$) exhibit 
a plateau of modes above some frequency scale $\omega^*\sim p^{-\delta}$ ($\sim z_c-z$, see below)
as proven by the collapse of the plateaus onsets in the right panels of Fig.\ref{f2}, and ($ii$)
display a minimum frequency $\omega_{\rm min}$, which does not scale with pressure
as $\omega^*$ does, as shown by the lack of data collapse of the low-frequency peak of $D(\omega)$ in the right panels of Fig.\ref{f2}. 
We indeed find that $\omega_{\rm min}\sim 1/\sqrt p$ for our constructed configurations shown in Fig.\ref{f3}{\bf a}, which is 
also true in flow as recalled in Fig.\ref{f3}{\bf c}.  
We note that the width of the low-frequency peak of $D(\omega)$ does not grow with increasing pressure, which indicates that the fluctuations in 
the spectra of ${\cal N}$ are indeed well-controlled. This is further supported by the limited spread of the clouds of points of Fig.\ref{f3},{\bf a} and {\bf c}, which represent
our entire data-set. 

Next, we test whether  our scheme correctly predicts the divergence of pressure (or viscosity) with coordination observed in flows. 
Fig.\ref{f3}{\bf b} shows the pressure $p$ vs. $\delta z \equiv z_c - z$ for each constructed configuration, 
and Fig.\ref{f3}{\bf d} represents the same quantity measured in flow. Remarkably, the scaling law $p\sim \delta z^{-1/\delta}$ with $\delta \approx 0.35$
holds for both ensembles of configurations. 

We now perform a scaling analysis of the properties of our  constructed configurations. We remove a fraction $\delta z/z_c$ of the weakest contacts (i.e. the smallest components of $|\tilde f_0\rangle$) and consider the operator ${\cal N}$  associated with the constructed configuration. 
We start by deriving a upper bound for the minimal eigenvalue $\omega_{\rm min}^2$ of ${\cal N}$, 
and an lower bound for the viscosity. ${\cal N}$ is symmetric hence
$\omega_{\rm min}^2\leq \langle x|{\cal N}|x\rangle$ for any
normalized vector $|x\rangle$. We consider the vector $|f_0\rangle$, the projection of $|\tilde f_0\rangle$
on the $N_c=(1-\delta z/z_c)\tilde N_c$  remaining contacts. Obviously lim$_{\delta z\rightarrow 0}||f_0||= ||\tilde f_0||\equiv 1$, thus:
\begin{equation}
\label{4}
\omega_{\rm min}^2\!\leq\!\frac{\langle f_0|{\cal N}|f_0\rangle}{||f_0||^2}\!\approx\!\langle f_0|{\cal N}|f_0\rangle \!\equiv\!
||{\cal T} | f_0\rangle||^2\!= \!||F_0||^2,
\end{equation}
where $||F_0||$ is the norm of the unbalanced force field associated with the contact force $|f_0\rangle$.
Removed contacts from a jammed configuration leads to unbalanced forces on particles which lost one  or more contacts. 
In particular, when a  contact $\alpha$ with a force $f_\alpha$ is removed,  each of the two adjacent particles carries an unbalanced
total force of amplitude $f_\alpha$.   In the limit $\delta z<<1$ most particles, whose total force is 
unbalanced, have only lost one contact. Thus:
\begin{equation}
\label{5}
||F_0||^2\approx 2 \sum_{\alpha_ r} f_{\alpha _r}^2=2 N_c\int_0^{f^*}P(f)f^2  df,
\end{equation}
where the sum is taken  over all the removed contacts $\alpha_ r$. Here $P(f)$ is the distribution
of the components of the vector $|\tilde f_0\rangle$ and $f^*$ is defined as $\int_0^{f^*}P(f)  df=\delta z/z_c$.
As Fig.\ref{f1} indicates, we observe that at low forces, 
$P(f/\langle f\rangle)\sim  (f\sqrt N_c)^\theta$, where the factor 
$1/\sqrt{N_c} \sim \langle f\rangle$ stems from the normalization of the vector $|\tilde f_0\rangle$.
We thus obtain two of our main predictions:
\begin{eqnarray}
\label{6}
\frac{f^*}{\langle f\rangle}&\sim&{\delta z}^{\frac{1}{1+\theta}},\\
\label{7}
\omega_{\rm min}^2&\leq& \delta z^\frac{3+\theta}{1+\theta}.
\end{eqnarray}

\begin{figure}
\onefigure[width=0.48\textwidth]{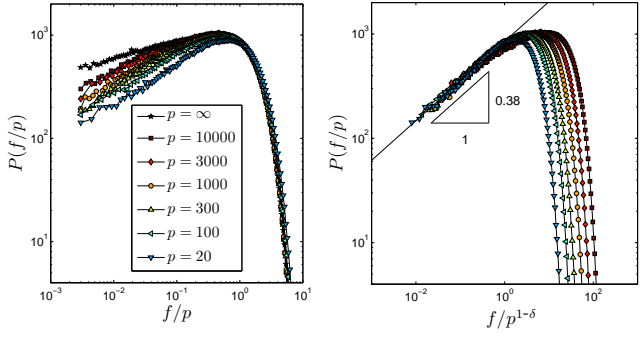}
\caption{{\bf a)} Distributions $P(f/p)$ of contact forces $f$ rescaled by the pressure $p$ measured in flow and shear-jammed configurations.
Measurements in flow are done at various $p$, as indicated in the legend. 
$P(f/p)$ erodes at low $f/p$ as the pressure diminishes away from the jamming threshold. 
{\bf b)} Distributions of contact forces vs.~rescaled force $f/(p\delta z) \sim f/p^{1-\delta}$.
This rescaling leads to the collapse of the  distribution at small force, indicating the presence of a scaling law.}
\label{f4}
\end{figure}

Eq.(\ref{6}) predicts the emergence of a characteristic force scale in flows, that vanishes
in relative terms as jamming is approached. In order to test this prediction we measure in flow the distributions
of contact forces for various pressures, see  Fig.\ref{f4}a. We observe an erosion in the distribution $P(f/p)$ of relative contact forces $f/p$, with a characteristic relative force that indeed decays near jamming. 
To  probe the scaling of this characteristic force, we seek a rescaling of the  contact force $f/f^*$ by some  $f^*$ that collapses the low-force tail of the distribution. 
The best collapse is found for $f^*\sim p \delta z \sim p^{1-\delta}$. As indicated in Table 1, 
this finding corresponds to our prediction for $\theta = 0$, and is still very close to our prediction using $\theta=0.18$. 
We note the difficulty in extracting the force scale $f^*$, as the crossover of the distributions of forces towards
the eroded regime is rather weak. Interestingly, we find that the rescaled distributions $P(f/f^*)$ scale  as $(f/f^*)^\chi$ with an exponent
$\chi\approx 0.38$.

We now test the inequality (\ref{7}). Assuming that this upper  bound is saturated, as excepted if the present variational argument captures the essential physics,  leads to a prediction for the scaling relation between $\omega_{\rm min}$ and $\delta z$. As shown in Table 1,  this prediction is in very good agreement with our observations. 

As noted above, $D(\omega)$ also displays a frequency scale $\omega^*\sim \delta z>>\omega_{\rm min}$ above which a 
plateau of modes appears. 
For completeness, we comment on $D(\omega)$ in the frequency range $[\omega_{\rm min},\omega^*]$.
Although it cannot be observed in our numerics due to the limited size of our systems, 
normal modes must be present in this interval.
Indeed, a local operator like ${\cal N}$ cannot have a single lowest eigenvalue 
$\omega_{\rm min}^2$ separated from the rest of the spectrum. This can be seen as follows:
if $|f_{\rm min}\rangle$ is the eigenvector corresponding to the eigenvalue $\omega_{\rm min}^2$, consider the 
family of eigenvectors $|f_{\vec q}\rangle$ build by modulating $|f_{\rm min}\rangle$ by 
plane waves:  $f_{{\vec q},\alpha}=\exp(i{\vec q} \cdot {\vec \alpha}) f_{\rm min,\alpha}$,
where ${\vec \alpha}$ is the position of contact $\alpha$.  The set of vectors
$|f_{\vec q}\rangle$  are approximatively orthonormal 
$\langle f_{\vec q'}|f_{\vec q}\rangle\approx \delta_{{\vec q},\vec{q'}}$.  
Furthermore, for small wave vectors the force balance that was nearly satisfied
in $|f_{\rm min}\rangle$ is only weakly perturbed, and one finds after a simple calculation that 
$\omega_{\vec q}^2\equiv \langle f_{\vec q}| {\cal N}|  f_{\vec q}\rangle= \omega_{\rm min}^2 + B q^2$,
where $B$ is a constant of order one. Using that the density of wave-vectors grows as $D(q)\sim q^{D-1}$, the  density of states of the frequencies 
$\omega_{\vec q}$ must then grow as
$D(\omega)\propto \omega(\omega^2-\omega_{\rm min}^2)^{(D-2)/2}$. 
Although the $\omega_{\vec q}^2$ are not exact eigenvalues of ${\cal N}$, 
we expect our estimate for $D(\omega)$ to be qualitatively correct for 
$\omega\in[\omega_{\rm min},\omega^*]$.

We now derive the divergence of the viscosity $\eta$ (or equivalently $p$ since $\eta\sim p$) with coordination. Using the convexity of the inverse function 
and Eq.(\ref{6}) leads to:
\begin{equation}
\label{8}
\langle f_0|{\cal N}^{-1}|f_0\rangle\geq\frac{1}{\langle f_0|{\cal N}|f_0\rangle} \sim \delta z^{-\frac{3+\theta}{1+\theta}}
\end{equation}
The finite friction coefficient $\mu^*$  in shear-jammed configurations implies that contact forces have a finite projection onto shear:
\begin{equation}
\label{3bis}
\hbox{lim}_{\delta z \rightarrow 0}\langle\gamma| f_0\rangle/||\gamma||=\langle\gamma|\tilde f_0\rangle/||\gamma||\equiv C>0.
\end{equation}
Together with Eqs.(\ref{1},\ref{8}), Eq.(\ref{3bis}) leads to:
\begin{equation}
\label{9}
\frac{\eta}{\xi_0}\!=\!\frac{\langle \gamma|{\cal N}^{-1} |\gamma\rangle}{\Omega}\!\approx\!\frac{\langle\gamma| f_0\rangle^2}{\Omega}\langle f_0|{\cal N}^{-1}|f_0\rangle\geq C^2 \frac{||\gamma||^2}{\Omega}\delta z^{-\frac{3+\theta}{1+\theta}}
\end{equation}
where we have neglected the contribution to the viscosity stemming from the
components of $|\gamma\rangle$ orthogonal to $|f_0\rangle$, which we expect to be small. 
From the definition of $|\gamma\rangle$ it is clear that $||\gamma||^2\sim d^2 N_c$, 
whereas $\Omega\sim N_c d^D$. Eq.(\ref{9}) thus yields the following lower bound for the rescaled 
viscosity  $\eta_r\equiv \eta/(d^{D-2}\xi_0)$:
\begin{equation}
\label{10}
\eta_r\geq A \delta z^{-\frac{3+\theta}{1+\theta}}
\end{equation}
where $A$ is a constant. Using that $p\sim\eta_r$ we can directly compare this prediction with the observations of Fig.(\ref{f3}). 
Table 1 shows that the observed scaling law is well-captured by  the saturation of Eq.(\ref{10}).
\begin{table}[h!]
\begin{center}
\begin{tabular}{|c | c | c | c|}
\hline
& \small{prediction} & \small{prediction} &  \\
\small{scaling}  & \small{$\theta\!=\! 0.18$} & \small{$\theta\!=\!0$} & \small{observation}  \\
\hline
$f^*\!/p \sim \delta z^{\frac{1}{1+\theta}}$& 0.84 & 1 & 1 \\
$\omega_{\rm min}^2\! \sim \!\delta z^{\frac{3+\theta}{1+\theta}}$ & 2.69 & 3 & 2.85 \\
$\eta\sim \delta z^{-\frac{3+\theta}{1+\theta}}$ & -2.69 & -3 & -2.85 \\
\hline
\end{tabular}
\caption{Predicted scaling exponents and exponents measured numerically in flow.
 Comparisons are made using the best fitted exponent $\theta = 0.18$ of the force distribution $P(f)$ of jammed states shown in Fig.\ref{f1}, and $\theta = 0$ assuming that the decay in $P(f)$ is logarithmic.}
\end{center}
\end{table}

\section{Conclusion}
In flow, the contact network constantly evolves, or rewires, via the formation and opening of contacts.  We have shown that a key aspect of this  process
is the negative selection of contacts that weakly affect  flow. Taking this effect into account enables one to derive three scaling relations between four exponents. These relations connect  the divergence of the viscosity  and spectral properties of dense flows
to two microscopic quantities: the coordination $z$ and the exponent $\theta$, which characterizes the density of weak contact forces in jammed configurations. 
Thus to obtain a complete description of the rheology, future works should compute the value of $\theta$ - see \cite{marginal} for recent results in this direction- and  the relation between coordination and packing fraction $z(\phi)$.   
In packings of soft repulsive particles with $\phi>\phi_c$, the coordination is the minimal one that guarantees mechanical stability  \cite{Wyart052,Wyart053}, a condition that determines $z(\phi)$. 
The concept of stability is not applicable for fluids however, and computing $z(\phi)$ in flow will presumably require a detailed description of the rewiring dynamics.

\acknowledgments
This work has been supported by the Sloan Fellowship, NSF DMR-1105387, the MRSEC program of the NSF DMR-0820341 and Petroleum 
Research Fund \#52031-DNI9.
We thank  A.~Grosberg,  E.~Vanden-Eijnden and D.~Kraft for comments on the manuscript. 


\begin{thebibliography}{0}

\bibitem{Einstein2}
  \Name{Einstein A.}
  \REVIEW{Annalen der Physik}{17}{1905}{549}.
  
\bibitem{Batchelor}
	\Name{Batchelor G.}
	\REVIEW{Journal of Fluid Mechanics }{83}{1977}{97}.

\bibitem{pouliquen2004}
	\Name{Pouliquen O.}
	\REVIEW{Phys. Rev. Lett.}{93}{2004}{248001}.

\bibitem{olsson}
	\Name{Olsson P., \and Teitel S.}
	\REVIEW{Phys.~Rev.~Lett.}{99}{2007}{178001}.

\bibitem{durian}
  \Name{Nordstrom K.~N., Verneuil E., Arratia P.~E., Basu A.,
Zhang Z., Yodh A.~G., Gollub J.~P., \and Durian D.~J.}
  \REVIEW{Phys.~Rev.~Lett.}{105}{2010}{175701}.

\bibitem{heussinger2010a}
  \Name{Heussinger C., Berthier L., \and Barrat J.~L.}
  \REVIEW{Europhys.~Lett.}{90}{2010}{20005}.
  
\bibitem{boyer}
	\Name{Boyer F., Guazzelli E., \and Pouliquen O.}
	\REVIEW{Phys.~Rev.~Lett.}{107}{2011}{188301}.

\bibitem{lespiat}
	\Name{Lespiat R., Cohen-Addad S., \and H{\"o}hler R.}
	\REVIEW{Phys. Rev. Lett.}{106}{2011}{148302}.
	
\bibitem{lemaitreroux}
	\Name{Lema{\^i}tre A., Roux J.-N., \and Chevoir F.}
	\REVIEW{Rheologica Acta}{48}{2009}{925}.

\bibitem{dacruz}
	\Name{da Cruz F., Emam S., Prochnow M., Roux J.-N., \and Chevoir F.}
	\REVIEW{Phys.~Rev.~E}{72}{2005}{021309}.

\bibitem{jop}
	\Name{Jop P., Pouliquen O., \and Forterre Y.}
	\REVIEW{Nature}{441}{2006}{727}.

\bibitem{staron}
	\Name{Staron L., Lagree P.-Y., Josserand C., \and Lhuillier D.}
	\REVIEW{Physics of fluids}{22}{2010}{113303}.

\bibitem{teitel2011}
	\Name{Olsson P. \and Teitel S.}
	\REVIEW{Phys.~Rev.~E}{83}{2011}{030302}.

\bibitem{heussinger2009}
	\Name{Heussinger C. \and Barrat J.-L.}
	\REVIEW{Phys.~Rev.~Lett.}{102}{2009}{218303}.

\bibitem{hatano08a}
	\Name{Hatano T.}
	\REVIEW{Phys.~Rev.~E}{79}{2008}{050301(R)}.

\bibitem{edan}
	\Name{Lerner E., D{\"u}ring G., \and Wyart M.}
	\REVIEW{Proc.~Natl.~Acad.~Sci}{109}{2012}{4798}.
	
\bibitem{footnote}
We account for finite-size corrections in $z_c$, see 
\emph{e.g.}~Goodrich C.~P., Liu A.~J., and Nagel S.~R., Phys.~Rev.~Lett.~{\bf 109}, (2012) 095704.


\bibitem{peyneau}
	\Name{Peyneau P.-E. \and Roux J.-N.}
	\REVIEW{Phys.~Rev.~E}{78}{2008}{041307}.
	
\bibitem{xu2006}
	\Name{N. Xu \and C. S. O'Hern}
	\REVIEW{Phys.~Rev.~E}{73}{2006}{061303}.

\bibitem{calladine}
	\Name{Calladine C.R.}
	\REVIEW{Int.~J.~Solids Struct.}{14}{1978}{161}.

\bibitem{simulation_paper}
	\Name{Lerner E., D{\"u}ring G., \and Wyart M.}
	\REVIEW{arXiv:}{1111.7225}{2011}{}.

\bibitem{ATbook}
	\Name{Allen M.~P. \and Tildesley D.~J.}
	\Book{Computer Simulations of Liquids}
	\Publ{Oxford Univ.~Press, New York}
	\Year{1991}.

\bibitem{selfAveragingMu}
	\Name{Peyneau P.-E. \and Roux J.-N.}
	\REVIEW{Phys.~Rev.~E}{78}{2008}{011307}.

\bibitem{marginal}
	\Name{Wyart M.}
	\REVIEW{Stability at Random Close Packing, \emph{to be published in} Phys.~Rev.~Lett.}{also \emph{arXiv:1202.0259}}{2012}

\bibitem{Wyart052}
	\Name{Wyart M., Silbert L. E., Nagel S. R., \and Witten T. A.}
	\REVIEW{Phys.~Rev.~E}{72}{2005}{051306}.

\bibitem{Wyart053}
	\Name{Wyart M.}
	\REVIEW{Annales de Phys}{30 (3)}{2005}{1}.

\end{thebibliography}

\end{document}